\documentclass[letter,twocolumn,seceq]{jpsj2_cond_mat} 
\title{First-Order Transition to Incommensurate Phase with Broken Lattice Rotation Symmetry in Frustrated Heisenberg Model}

\author{Ryo Tamura and Naoki Kawashima}

\inst{Institute for Solid State Physics, University of Tokyo, 5-1-5 Kashiwa-no-ha, Kashiwa, Chiba 277-8581, Japan}

\abst{We study a finite-temperature phase transition in the two-dimensional classical Heisenberg model on a triangular lattice
with a ferromagnetic nearest-neighbor interaction $J_1$ and an antiferromagnetic third-nearest-neighbor interaction $J_3$
using a Monte Carlo method.
Apart from a trivial degeneracy corresponding to O(3) spin rotations,
the ground state for $J_3 \neq 0$ has a threefold degeneracy corresponding to 120 degree lattice rotations.
We find that this model exhibits a first-order phase transition with the breaking of the threefold symmetry 
when the interaction ratio is $J_3/J_1=-3$.}

\kword{incommensurate phase, finite temperature transition, frustration, $\text{NiGa}_2\text{S}_4$}

\begin{document}
\maketitle


The magnetic compound $\text{NiGa}_2\text{S}_4$ is a rare example of a two-dimensional triangular lattice antiferromagnet 
and stays in a spin-disordered state at low temperatures.
Interesting behaviors of this material have been observed by Nakatsuji and coworkers \cite{nakatsuji1,nambu,nakatsuji2,nakatsuji3};
the magnetic part of the specific heat exhibits an unusual double-peak structure
and the magnetic susceptibility gradually increases without divergent behavior as the temperature decreases.
They have estimated the Weiss temperature $\theta_{W} \cong -80$ K from the susceptibility results,
where this temperature corresponds to the high-temperature peak observed in the specific heat.
Below the low-temperature peak, the specific heat shows a $T^2$ dependence
and a short-range noncollinear order develops.
Indeed, the development of incommensurate (IC) spin correlation has been confirmed by magnetic neutron scattering experiment.
This observation was explained by the mean-field calculation of a Heisenberg model 
with a dominant antiferromagnetic (AF) third-nearest-neighbor interaction.

The antiferromagnetic Heisenberg model on the triangular lattice has been studied exhaustively.
Kawamura and Miyashita argued \cite{kawamura1} that a classical model with only an AF nearest-neighbor (NN) interaction exhibits
a topological phase transition driven by the dissociation of vortices at a finite temperature.
One of the theoretical scenarios \cite{kawamura2} proposed for $\text{NiGa}_2\text{S}_4$ is based on this mechanism,
taking into account the effect of biquadratic interactions.
Another scenario \cite{tsunetsugu,lauchli,bhattacharjee} is the spin nematic ordering based on the $S=1$ spin model 
with NN bilinear-biquadratic interactions.
Although these theoretical works predicted a number of interesting physical phenomena, 
they do not fully take into account the characteristic spatial structure of $\text{NiGa}_2\text{S}_4$,
that causes the IC phase.
The IC phase was observed in the model with the AF NN and AF second-NN interactions \cite{jolicoeur}.
If the second-NN interaction is dominant, there are three IC structures of the ground state
that can be transformed to each other by 120 degree lattice rotations.
It is easy to see that a similar threefold degeneracy exists when the third-NN interaction is dominant.
Because of this discrete symmetry, a finite temperature phase transition can take place without violating the Mermin and Wagner's theorem \cite{mermin}.
Therefore, in order to understand finite temperature properties of $\text{NiGa}_2\text{S}_4$,
it is necessary to discuss the possibility of such a phase transition.


In the present study, we investigate a two-dimensional classical Heisenberg model on a triangular lattice
with a ferromagnetic NN interaction $J_1$ ($<0$) and an antiferromagnetic third-NN interaction $J_3$ ($>0$).
The model Hamiltonian is given by
\begin{align}
\mathcal{H} = J_1 \sum_{{\langle i , j \rangle}_{\text{NN}}} \boldsymbol{s}_i \cdot \boldsymbol{s}_j +J_3 \sum_{{\langle i , j \rangle}_{\text{3rd.NN}}} \boldsymbol{s}_i \cdot \boldsymbol{s}_j \label{eq:hamiltonian},
\end{align}
where $\boldsymbol{s}_i$ is the vector spin of unit length.
The first sum runs over NN pairs of sites and the second sum runs over third-NN pairs.
Here, we apply the periodic boundary condition and assume that $J_3$ is dominant ($|J_3| > |J_1|$).
Indeed, the existence of the dominant interaction $J_3$ is supported 
not only by photoemission spectroscopy \cite{takubo} but also by first-principles calculation \cite{mazin}.
We neglect the second-NN interaction $J_2$, 
because the magnetic neutron scattering result suggests that $|J_2|$ is much smaller than $|J_1|$ or $|J_3|$ \cite{nakatsuji1}.
In addition, the above-mentioned threefold degenerate structure of the ground state 
does not change even if we introduce a weak second-NN interaction.
Therefore, we can neglect $J_2$ without changing essential physics.

\begin{figure}
\includegraphics[trim=-10mm 0mm 0mm 0mm ,scale=0.40]{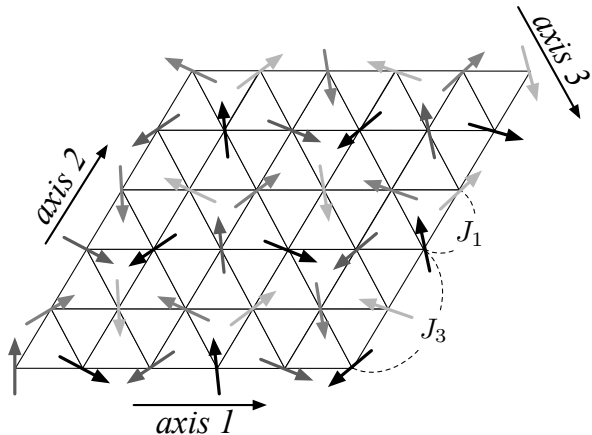} 
\caption{\label{fig:15738Fig1} Schematic picture of spin configuration at $\boldsymbol{k}=\pm (k,0)$.
Axis 1 is characterized by $k$, and axes 2 and 3 by $k/2$.
There is a threefold degeneracy corresponding to 120 degree lattice rotations.
The spin configuration on each of the four sublattices is close to, but not exactly, the 120 degree structure.}
\end{figure}

We first study the ground-state classical spin configuration of the Hamiltonian eq. (\ref{eq:hamiltonian}).
If there is no NN interaction $J_1$,
the ground-state spin configuration on each of the four sublattices, which has a double period of the lattice, 
is a 120 degree structure.
Since $J_1$ generates correlation between sublattices,
the spin configuration of each sublattice is distorted from the 120 degree structure.
Hence, the ground state becomes the IC state.
This IC phase is described by a spiral configuration with the wave vector $\boldsymbol{k}$ that minimizes the Fourier transform
of the interactions.
There are six such wave vectors $\boldsymbol{k}$ inside the first Brillouin zone of the triangular lattice.
The spiral configuration is given by
\begin{align}
\boldsymbol{s}_i = \boldsymbol{R} \cos \boldsymbol{k} \cdot \boldsymbol{r}_i - \boldsymbol{I} \sin \boldsymbol{k} \cdot \boldsymbol{r}_i,
\end{align}
where $\boldsymbol{R}$ and $\boldsymbol{I}$ are two arbitrary orthogonal unit vectors 
and $\boldsymbol{r}_i$ is the position of site $i$ in the real space on the triangular lattice.
In the case of Heisenberg spins,
the difference between $\boldsymbol{k}$ and $-\boldsymbol{k}$ can be absorbed in the definitions of $\boldsymbol{R}$ and $\boldsymbol{I}$. 
Thus, there are three distinct groups of states corresponding to 
$\boldsymbol{k} = \pm (k,0)$, $\pm (\frac{1}{2} k,\frac{\sqrt{3}}{2} k)$, and $\pm (\frac{1}{2} k,- \frac{\sqrt{3}}{2} k)$
in the IC phase.
The schematic picture of the spin configuration at $\boldsymbol{k}=\pm (k,0)$ is shown in Fig. \ref{fig:15738Fig1}.
The spiral spin configuration along one of the three axes can be characterized by the wave number $k$,
while others by $k/2$.
These three state groups are well separated only for IC wave numbers.
In other words, if the ordering is commensurate, i.e., if the ground state is a 60 degree structure,
the 120 degree spatial rotation can be achieved as a result of continuous spin rotation with the total energy being fixed.
However, a technical difficulty characteristic of the IC nature of the ground state arises from the fact that
the wave vector cannot take an arbitrary value in the reciprocal space when the system is finite.
Thus, the wave vector $\boldsymbol{k}$ of the ground state depends on the system size in an irregular fashion.
For the interaction ratio $J_3/J_1$, we use the value $-3$ in the following calculations.
According to the original experiment \cite{nakatsuji1}, $J_3/J_1=-3 \sim -10$.
Therefore, the present value is the lowest, which is consistent with the experimental result.
We choose the smallest value for a technical reason.
For larger values, relatively large systems are required for stabilizing the structure 
that converges to the true IC ground state in the thermodynamic limit.
A more systematic study for various $J_3/J_1$ values will be published elsewhere \cite{tamura}.
In this case, the wave number of the IC spin configuration is $k=|\boldsymbol{k}| \cong 1.92188$
in the thermodynamic limit.
These results indicate the possibility of the spontaneous breaking of the threefold symmetry at a finite temperature.

We perform classical Monte Carlo simulations based on the standard heat-bath method.
Each run contains $2\sim10\times10^6$ Monte Carlo steps per spin at each temperature.
We make $40\sim64$ independent runs for each size to evaluate the statistical errors.
Throughout this letter, the Boltzmann constant is set to unity.

\begin{figure}
\includegraphics[trim=0mm 15mm 0mm 5mm ,scale=0.43]{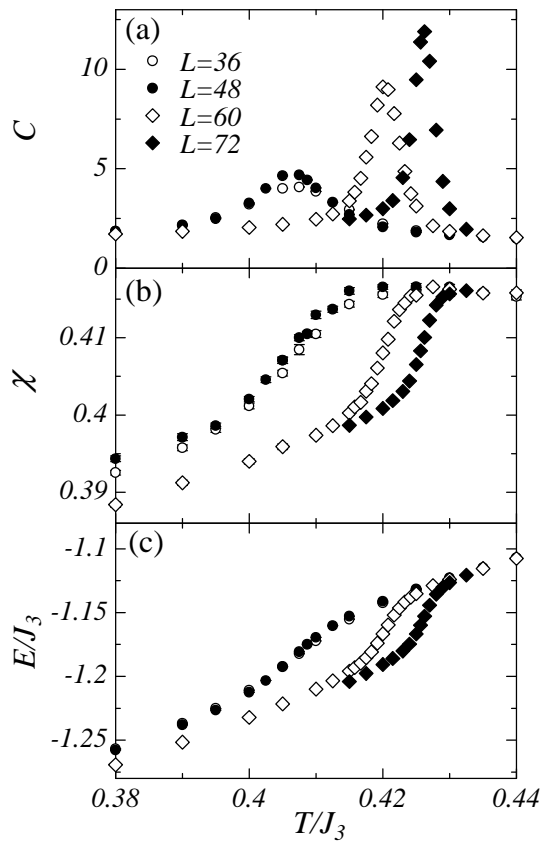} 
\caption{\label{fig:15738Fig2} Temperature dependences of the (a) specific heat $C$, (b) uniform susceptibility $\chi$, and (c) internal energy $E$.}
\end{figure}

To discuss a finite-temperature phase transition,
we calculate the specific heat $C$, the uniform susceptibility $\chi$, and the internal energy $E$.
$C$ and $\chi$ are defined by
\begin{align}
C &= \frac{1}{L^2} \frac{\langle E^2 \rangle - \langle E \rangle^2}{T^2}, \\
\chi &= \frac{1}{L^2} \frac{{\langle \boldsymbol{M}^2 \rangle }}{T/J_3}, \\
\boldsymbol{M} &= \left( \sum_{\text{site}} s_x , \sum_{\text{site}} s_y , \sum_{\text{site}} s_z \right), 
\end{align}
where $L$ is the system size and $\langle \cdots \rangle$ indicates the thermal average.
We show the results in Fig. \ref{fig:15738Fig2}.
The specific heat exhibits a single peak,
which is narrower and higher for larger systems, indicating a phase transition.
As mentioned above, the peak position depends on the system size irregularly owing to the compatibility 
of the ordering wave vector with the system size.
Therefore, we cannot reliably estimate the transition temperature.
We introduce a characteristic temperature $T_c (L)$ depending on the lattice size,
which is defined as the peak position of the specific heat.
The uniform susceptibility and internal energy decrease rapidly at around $T_c (L)$.
Clearly, the phase transition of the present model is different from 
that of the AF Heisenberg model on the triangular lattice with no third-NN interactions \cite{kawamura1}.
 
\begin{figure}
\includegraphics[trim=0mm 115mm 0mm 10mm ,scale=0.40]{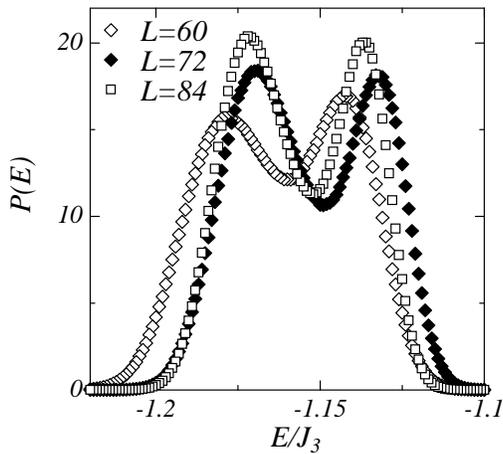} 
\caption{\label{fig:15738Fig3} Energy distributions $P(E)$ near $T_c (L)$
for $L=60 \ (T/J_3=0.4205)$, $72 \ (T/J_3=0.4262)$, and $84 \ (T/J_3=0.4240)$.}
\end{figure}

While Fig. \ref{fig:15738Fig2} indicates the phase transition, we study the energy distribution to find the order of this transition.
If the system is at the first-order transition temperature, the energy distribution $P (E)$ should be bimodal.
The distribution $P(E)$ at $T \cong T_c (L)$ is shown in Fig. \ref{fig:15738Fig3}.
Again, we cannot avoid the irregularity due to the incommensurability and finiteness of the system. 
However, we can at least see a clear evidence of the bimodal energy distribution,
i.e., the valley in the middle of the distribution deepens with increasing lattice size.
Therefore, we consider that the phase transition is of the first order.
The same results have been obtained irrespective of whether the initial spin configuration is set to that of the IC order 
or a random spin configuration. 
Although not shown in Fig. \ref{fig:15738Fig3}, the distribution $P(E)$ is singly peaked for $L=36$ and $48$.
This indicates that it takes considerably large systems to observe the first sign of the first-order transition.

\begin{figure}
\includegraphics[trim=-5mm 80mm 0mm 10mm ,scale=0.43]{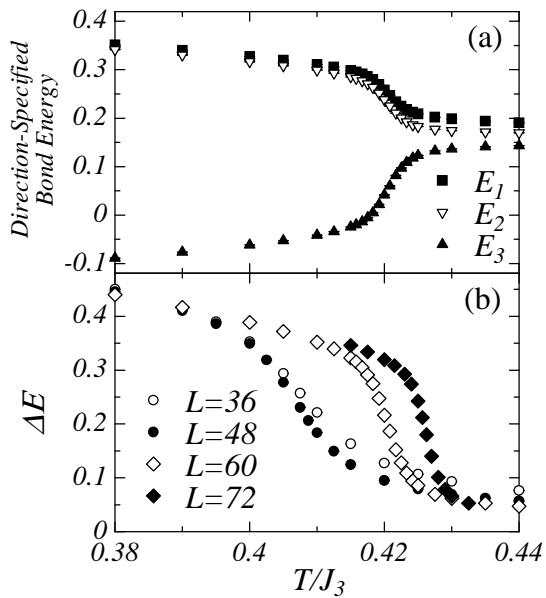} 
\caption{\label{fig:15738Fig4} Temperature dependences of the (a) direction-specified bond energies $E_1$, $E_2$, and $E_3$ for $L=60$ 
and (b) $\Delta E \ (=E_1-E_3)$ for various sizes.}
\end{figure}

From the ground state properties of the present model,
it is natural to expect the spontaneous breaking of the threefold symmetry.
Based on this expectation,
we calculate average bond energies along the three axes separately.
To be specific,
\begin{align}
\varepsilon_{\mu} &= \frac{1}{L^2} \sum_{{\langle i , j \rangle}_{\text{NN}} \ \parallel \ \text{axis} \ \mu} \boldsymbol{s}_i \cdot \boldsymbol{s}_j \ \ \ \ (\mu = 1,2,3).
\end{align}
We sort the three averages in descending order and define $E_1$, $E_2$, and $E_3$ as
\begin{align}
E_1 &= \langle \text{max} \{\varepsilon_1,\varepsilon_2,\varepsilon_3 \} \rangle , \notag \\
E_2 &= \langle \text{mid} \{\varepsilon_1,\varepsilon_2,\varepsilon_3 \} \rangle, \ \text{and} \\
E_3 &= \langle \text{min} \{\varepsilon_1,\varepsilon_2,\varepsilon_3 \} \rangle. \notag
\end{align}
In Fig. \ref{fig:15738Fig4}(a), we present the temperature dependence of the average direction-specified bond energies for $L=60$.
$E_1$ and $E_2$ increase but $E_3$ decreases below $T_c (L)$.
This result implies that the threefold symmetry is broken and one of the three axes is selected.
The existence of such a characteristic axis can also be seen in a snap shot of the spin configuration.
To study the anomalous behavior of the average direction-specified bond energies more quantitatively, 
we estimate the energy difference defined by $\Delta E=E_1 - E_3$.
The temperature dependence of $\Delta E$ is shown in Fig. \ref{fig:15738Fig4}(b).
$\Delta E$ abruptly increases around $T_c (L)$ and the slope of $\Delta E$ seems to diverge as the lattice size increases.
From these results, we conclude that the first-order phase transition is accompanied by the spontaneous breaking of the threefold symmetry.

\begin{figure}
\includegraphics[trim=-5mm 70mm 0mm 20mm ,scale=0.43]{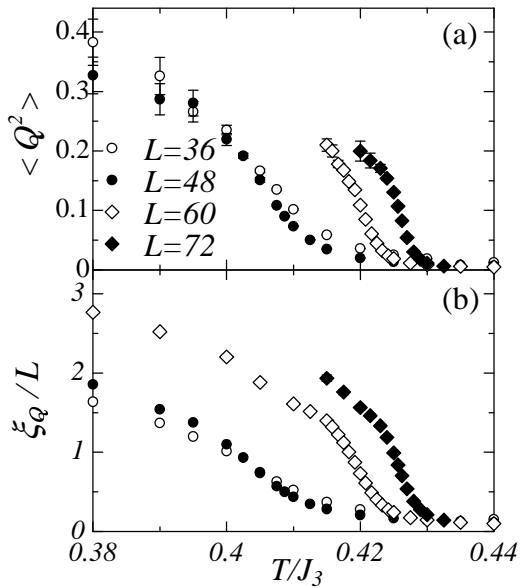} 
\caption{\label{fig:15738Fig5} Temperature dependences of the (a) $\langle Q^2 \rangle$ and (b) correlation length $\xi_Q/L$.}
\end{figure}

In the two-dimensional classical Heisenberg spin system,
the true long-range order of spins is prohibited by the Mermin and Wagner's theorem \cite{mermin}.
Therefore, the occurrence of the spontaneous symmetry breaking must be attributed to an order parameter other than bare spins.
We consider the order parameter 
\begin{align}
Q &= \frac{1}{L^2} \sum_{\boldsymbol{r} \in \triangle} Q (\boldsymbol{r}), \label{eq:Q} \\
Q(\boldsymbol{r}) &= \boldsymbol{s}_1 \cdot \boldsymbol{s}_2 + \boldsymbol{s}_2 \cdot \boldsymbol{s}_3 - 2 \boldsymbol{s}_3 \cdot \boldsymbol{s}_1, \label{eq:Q(R)} 
\end{align}
where $Q (\boldsymbol{r})$ is defined on an upward elementary triangle.
The corner sites 1, 2, and 3 are numbered counterclockwise for a triangle.
The sum of eq. (\ref{eq:Q}) runs over all the upward elementary triangles.
The quantity $Q$ takes a finite value in the ordered phase and 0 in the disordered phase.
The temperature dependence of $\langle Q^2 \rangle$ is shown in Fig. \ref{fig:15738Fig5}(a).
In the disordered phase, the correlation length $\xi_Q$ of $Q(\boldsymbol{r})$ can be obtained from the structure factor  $S_Q (\boldsymbol{k}_0)$
using the Ornstein-Zernike form \cite{southern}
\begin{align}
\xi_Q &= \frac{1}{|\boldsymbol{k}_0|} \sqrt{\frac{S_Q (\boldsymbol{0})}{S_Q (\boldsymbol{k}_0)}-1}, \label{eq:xi} \\
S_Q (\boldsymbol{k}_0) &= \frac{1}{L^2} \sum_{i,j} \langle Q (\boldsymbol{r}_i) Q (\boldsymbol{r}_j) \rangle 
e^{i \boldsymbol{k}_0 \cdot (\boldsymbol{r}_i-\boldsymbol{r}_j)},
\end{align}
where \textbf{0} is the ordering vector of $Q (\boldsymbol{r})$ and $\boldsymbol{k}_0$ is any wave vector close to the ordering vector.
Note that eq. (\ref{eq:xi}) does not yield the correct values below $T_c (L)$.
In Fig. \ref{fig:15738Fig5}(b), we show the temperature dependence of the correlation length.
The correlation length discontinuously increases and becomes larger than the lattice size at around $T_c (L)$.
Thus, the $Q (\boldsymbol{r})$ degree of freedom freezes at $T_c (L)$.

We consider implications of the present results on the experimental results of $\text{NiGa}_2\text{S}_4$.
The important points in the experimental results are that the IC phase emerges at low temperatures
and the specific heat has a broad double-peak structure.
The present results are consistent with the incommensurability observed in the neutron scattering experiments.
On the other hand, the specific heat of the present model only exhibits a single divergent peak.
It corresponds to the low-temperature peak (around 10 K) observed in the experiment,
because the experimental estimate $J_3 \cong 30$ K \cite{nakatsuji4} 
is too small to be responsible for the high-temperature peak at around 80 K.
Considering the scenario based on the topological transition \cite{kawamura1},
we may observe two peaks, one corresponding to the first-order transition we discuss in the present article,
and the other corresponding to the topological transition.
However, we have not observed any sign of the topological transition in the present case of $J_3/J_1=-3$;
there is no clear sign for a first-order transition in the experiment,
and, as mentioned above, the higher transition temperature is too high.
Therefore, it is not likely that all experimental evidences can be explained by the present model.
The inclusion of various other terms, such as spin anisotropy \cite{miyashita} and magnetic field, \cite{kawamura3} will be discussed elsewhere \cite{tamura}.

It may be suitable to make a few comments on the first-order phase transition in the present model.
The three-state Potts model is the representative model with threefold symmetry breaking \cite{Wu}.
In two dimensions, this model has a second-order phase transition at a finite temperature,
in a strong contrast to the present model,
despite the same type of symmetry breaking.
This is not very surprising considering that the order of the phase transition is determined by physics on short-length scales.
In the present model, short-range physics is strongly affected by the IC nature of the magnetic ordering.
As we have discussed above, the IC ordering may be closely related to the first-order transition observed in the present study,
because the commensurate structure (i.e., the exactly 60 degree structure) cannot energetically separate the three degenerate thermodynamic states.


To summarize, we have studied the two-dimensional classical Heisenberg model on the triangular lattice
with the ferromagnetic nearest-neighbor interaction $J_1$ and the antiferromagnetic third-nearest-neighbor interaction $J_3$.
We have found that the finite temperature phase transition is of the first order,
and that this transition is accompanied by the breaking of the threefold symmetry.
The incommensurate nature of the ground state may be essential to the present type of phase transition.


\section*{Acknowledgment}

We would like to thank S. Nakatsuji, T. Suzuki, and H. Kawamura for useful discussions.
The present work is financially supported by MEXT Grant-in-Aid for Scientific Research (B) (19340109), 
MEXT Grant-in-Aid for Scientific Research on Priority Areas ``Novel States of Matter Induced by Frustration'' (19052004),
and Next Generation Supercomputing Project, Nanoscience Program, MEXT, Japan.
The computation in the present work was performed on computers at the Supercomputer
Center, Institute for Solid State Physics, University of Tokyo.


\end{document}